\documentclass[a4paper,11pt]{article}
\usepackage{graphicx}
\newcommand{\newc}{\newcommand}
\newc{\lra}{\leftrightarrow}
\newc{\beq}{\begin{equation}}
\newc{\eeq}{\end{equation}}
\newc{\barr}{\begin{eqnarray}}
\newc{\earr}{\end{eqnarray}}

\begin{document}
\bibliographystyle{plain}

%%%%%%% TITLE %%%%%%
\begin{flushleft}
\hspace{4.0cm}NEUTRINOS IN A SPHERICAL BOX
 \vspace{1cm}
\\
%%%%%%%%%%%%%%%%%%%%
%%%%% Author List %%%%%%%
\hspace{4.0cm} Giomataris$^{1}$ and J.D. Vergados$^{2}$
\\
%%%%%%%%%%%%%%%%%%%%%%%%%
%%%%% Affiliation %%%%%%%

1 CEA, Saclay, DAPNIA, Gif-sur-Yvette, Cedex,France\\
2 Theoretical Physics Division, University of Ioannina, Gr 451 10,
Ioannina, Greece
\\E-mail:Vergados@cc.uoi.gr
%%%%%%%%%%%%%%%%%%%%%%%%%%
\end{flushleft}
\vspace{0.5cm}
%%%%% Abstract %%%%%%%%%%%
\begin{abstract} \baselineskip 12pt
The purpose of the present paper is to study the neutrino
properties as they may appear in the low energy neutrinos emitted
in triton decay:
$$^3_1 H \rightarrow ^3_2He +e^-+\tilde{\nu}_e$$
with maximum neutrino energy of $18.6~KeV$. The technical
challenges to this end can be summarized as building a very large
TPC capable of detecting low energy recoils, down to a few 100 eV,
within the required low background constraints. More specifically
We propose the development of a spherical gaseous TPC of about
10-m in radius and a 200 Mcurie triton source in the center of
curvature. One can list a number of exciting studies, concerning
fundamental physics issues, that could be made using a large
volume TPC and low energy antineutrinos: 1) The oscillation length
involving the small angle $\delta=\sin{\theta_{13}}$, directly measured in our
$\nu_e$ disappearance experiment,   is fully contained inside the detector.
 Measuring the counting rate of neutrino-electron
elastic scattering as function of the distance of the source will
give a precise and unambiguous measurement of the oscillation
parameters free of systematic errors. In fact first estimations
show that even with a year's data taking a sensitivity of a few percent
 for the measurement of
the above angle will be achieved. 2) The low energy detection threshold offers a
unique sensitivity for the neutrino magnetic moment which is about
two orders of magnitude beyond the current experimental limit of
$10^{-10}\mu_B$. 3) Scattering at such low neutrino energies has
never been studied and any departure from the expected behavior
may be an indication of new physics beyond the standard model.
 We present a summary of various theoretical
studies and possible measurements, including a precise measurement of
the Weinberg angle at very low momentum transfer.
\end{abstract}
\section{Introduction.}
\label{secint}
Neutrinos are the only particles in nature, which are
characterized by weak interactions only. They are thus expected to
provide the laboratory for understanding the fundamental laws of
nature. Furthermore they are electrically neutral particles
characterized by a very small mass. Thus it is an open question
whether they are truly neutral, in which case the particle
coincides with its own antiparticle , i.e. they are Majorana
particles, or they are characterized by some charge, in which case
they are of the Dirac type, i.e the particle is different from
its antiparticle \cite{VERGADOS}. It is also expected that the
neutrinos produced in weak interactions are not eigenstates of the
world Hamiltonian, they are not stationary states, in which case
one expects them to exhibit oscillations \cite{VERGADOS,VOGBEAC} .
As a matter of fact such neutrino oscillations seem to have
observed in atmospheric neutrino \cite{SUPERKAMIOKANDE},
interpreted as
 $\nu_{\mu} \rightarrow \nu_{\tau}$ oscillations, as well as
 $\nu_e$ disappearance in solar neutrinos \cite{SOLAROSC}. These
 results have been recently confirmed by the KamLAND experiment \cite{KAMLAND},
 which exhibits evidence for reactor antineutrino disappearance.
 This has been followed by an avalanche of interesting analyses
 \cite{BAHCALL02}-\cite{BARGER02}.
 The purpose of the present paper is to discuss a new experiment to
study the above neutrino properties as they may appear in the low
energy neutrinos emitted in triton decay:
$$^3_1 H \rightarrow ^3_2He +e^-+\tilde{\nu}_e$$
with maximum neutrino energy of $18.6~KeV$. This process has
previously been suggested \cite{SHROCK} as a means of studying
heavy neutrinos like the now extinct $17 KeV$ neutrino. The
detection will be accomplished employing gaseous Micromegas,large
TPC (Time Projection Counters) detectors with good energy
resolution and low background \cite{GIOMATAR}. In addition in
this new experiment we hope to observe or set much more stringent
constraints on the neutrino magnetic moments. This very
interesting question has been around for a number of years and it
has been revived recently \cite{VogEng}-\cite{TROFIMOV}. The
existence of the neutrino magnetic moment can be demonstrated
either in neutrino oscillations in the presence of strong
magnetic fields or in electron neutrino scattering. The latter is
expected to dominate over the weak interaction in the triton
experiment since the energy of the outgoing electron is very
small. Furthermore the possibility of directional experiments
will provide additional interesting signatures. Even experiments
involving polarized electron targets are beginning to be
contemplated \cite{RASHBA}. There are a number of exciting
studies, of fundamental physics issues, that could be made using
a large volume TPC and low energy antineutrinos:
\begin{itemize}
\item The oscillation length is comparable to the length of the detector. Measuring
the counting rate of neutrino elastic scattering as function of
the distance of the source will give a precise and unambiguous
measurement of the oscillation parameters free of systematic
errors. First estimations show that a sensitivity of a few percent
for the measurement of $\sin^2{\theta_{13}}$.
\item The low energy detection threshold offers a unique sensitivity for the
neutrino magnetic moment, which is about two orders of magnitude
beyond the current experimental limit of $10^{-10}\mu_B$. In our
estimates below we will use the optimistic value of
$10^{-12}\mu_B$.
\item Scattering at such low neutrino energies has never been studied before. In
addition one may exploit the extra signature provided by the
photon in radiative electron neutrino scattering. As a result any
departure from the expected behavior may be  an indication of
physics beyond the standard model.
\end{itemize}
 In the following we will present a summary of various theoretical studies and
possible novel measurements
\section{Neutrino masses as extracted from various experiments}
\label{secnme}
 At this point it instructive to elaborate a little
bit on the neutrino mass combinations entering various
experiments.
\begin{itemize}
\item {\bf Cosmological Constraints}. We get \cite{WMAP,2dFGRS}:
$$\Omega_{\nu}~h^2\leq 0.0076(95\%~CL)$$
$$\frac{\Sigma_i~m_i}{93.5~eV}=\Omega_{\nu}~h^2\Rightarrow$$
$$\Sigma_i~m_i\leq 0.71~eV/c^2~~(95\%~CL)$$
(Majorana neutrinos).\\
 The limit becomes $1.05~eV$ without the Ly-$\alpha$ forest data
For Dirac neutrinos the value of the upper limit is half the
above.
\item Neutrino oscillations. \\
These in principle, determine the  mixing matrix and
   two independent mass-squared differences, e.g.
 $$\Delta m^2_{21} = m^2_2 - m^2_1~~,~~ \Delta m^2_{31} = m^2_3 - m^2_1$$
They cannot determine:
\begin{enumerate}
\item the scale of the masses, e.g. the lowest
eigenvalue $m_1$ and
\item the two relative Majorana phases.
\end{enumerate}
\item The end point triton decay.

This can determine one of the masses, e.g. $m_1$ by measuring:
\begin{equation}
(m_{\nu})_{1 \beta} \equiv m_{\nu}=| \sum_{j=1}^{3}
 U^*_{ej}U_{ej} m^2_j |^{1/2} \, , ~U=U^{11}
\label{mass.3}
\end{equation}
 Once $m_1$ is known one can find\\
  $$m_2=[\delta m^2_{21}+m^2_1]^{1/2}~~,~~m_3=[\delta m^2_{31}+m^2_1]^{1/2}$$
 provided, of course that the mixing matrix is known. \\
Since the Majorana phases do not appear, this experiment cannot
differentiate between Dirac and Majorana neutrinos. This can only
be done via lepton violating processes, like: \item $0\nu \beta
\beta $ decay.

This provides an additional   independent linear combinations of
the masses and the Majorana phases.
\begin{equation}
\langle m_{\nu}\rangle_{2 \beta} \equiv \langle m_{\nu}\rangle =
|\sum _{j=1}^{3} U_{ej}U_{ej}e^{i\lambda _j} m_j| \label{mass.1}
\end{equation}
\item and muon to positron conversion.

This also provides an additional relation
\begin{equation}
\langle m_{\nu}\rangle_{\mu e+}=|\sum_{j=1}^{3}
                   U^*_{\mu j}U^*_{ej}e^{-i\lambda _j}m_j| \, .
\label{mass.2}
\end{equation}
\end{itemize}
 Thus the two independent relative CP phases can in principle be measurable.
{\bf So these three types of experiments together can specify all
parameters not settled by the neutrino oscillation experiments}.

 Anyway from the neutrino oscillation data alone we cannot infer
 the mass scale. Thus the following scenarios emerge
 \begin{enumerate}
 \item the lightest neutrino is $m_1$ and its mass is very small.
 This is the normal hierarchy scenario. Then:
$$\Delta m^2_{21} = m^2_2~~,~~ \Delta m^2_{31} = m^2_3$$
\item The inverted hierarchy scenario. In this case the mass $m_3$
is very small. Then:
$$\Delta m^2_{21} = m^2_2 - m^2_1~~,~~ \Delta m^2_{31} = m^2_1$$
\item The degenerate scenario. In such a situation all masses are about
equal and much larger than the differences appearing in neutrino
oscillations. In this case we can obtain limits on the mass scale
as follows:
\begin {itemize}
\item From  triton decay. Then \cite{LOBASHEV}
$$ m_1 \approx (m_{\nu})_{1 \beta}\leq 2.2 eV$$
This limit is expected to substantially improve in the future
\cite{KATRIN}. \item From  $0\nu~\beta\beta$ decay. The analysis
now depends on the mixing matrix and the CP phases of the Majorana
neutrino eigenstates \cite{VERGADOS} (see discussion below). The
best limit coming from $0\nu~\beta\beta$ decay is:
 $$m_1 \approx\langle m_{\nu}\rangle_{2 \beta}\leq 0.5~eV,
m_1 \approx \frac{\langle m_{\nu}\rangle_{2 \beta}}{\cos{2
\theta_{solar}}}\approx 2 \langle m_{\nu}\rangle_{2 \beta}\leq
1.0~eV,$$ for  relative CP phase of the two strongly admixed
states is $0$ and $\pi$ respectively.

These limits are going to greatly improve in the next generation
of experiments, see e.g. the review by Vergados  \cite{VERGADOS}
and the experimental references therein.
\end{itemize}
\end{enumerate}

\section{Elastic electron neutrino scattering.}
\label{secens}

 The elastic neutrino electron scattering, which has played an
 important role in physics \cite{REINES}, is very crucial in our
 investigation, since it will be employed for the detection of
 neutrinos. So we will briefly discuss it before we embark on the
 discussion of the apparatus.

  Following the pioneering work of 't Hooft \cite{HOOFT} as well as
  the subsequent work of
Vogel and Engel \cite{VogEng} one can write the relevant
differential cross section as follows:
%\end{document}
\begin{equation}
\frac{d\sigma}{dT}=\left(\frac{d\sigma}{dT}\right)_{weak}+
\left(\frac{d\sigma}{dT}\right)_{EM} \label{elas1a}
\end{equation}
We ignored the contribution due to the neutrino charged radius. We
will not consider separately the scattering of electrons bound in
the atoms, since such effects have recently been found to be small
\cite{GOUNARIS}.

The cross section due to weak interaction alone takes the form
\cite{VogEng}:
\begin{eqnarray}
 \left(\frac{d\sigma}{dT}\right)_{weak}&=&\frac{G^2_F m_e}{2 \pi}
 [ (g_V+g_A)^2+ (g_V-g_A)^2 (1-\frac{T}{E_{\nu}})^2\\
\nonumber
&+& (g_A^2-g_V^2)\frac{m_eT}{E^2_{\nu}}      ]
 \label{elasw}
 \end{eqnarray}
 where
 $$g_V=2\sin^2\theta_W+1/2~~for~~ \nu_e~~~~,~~~~g_V=2\sin^2\theta_W-1/2~~for~~ \nu_{\mu},\nu_{\tau}$$
 $$g_A=1/2~~for~~ \nu_e~~~~,~~~~g_A=-1/2~~for~~ \nu_{\mu},\nu_{\tau}$$
 For antineutrinos $g_A\rightarrow-g_A$. To set the scale we see
 that
\beq \frac{G^2_F m_e}{2 \pi}=0.445\times 10^{-48}~\frac{m^2}{MeV}
\label{weekval} \eeq
 In the above expressions for the $\nu_{\mu},\nu_{\tau}$ only the
 neutral current has been included, while for $\nu_e$ both the
 neutral and the charged current contribute.

 The second piece of the cross-section becomes:
\begin{equation}
\left(\frac{d\sigma}{dT}\right)_{EM}= \pi (\frac{ \alpha}{m_e})^2
(\frac{ \mu_{l}}{\mu_B})^2 \frac{1}{T} \left(1-\frac{T}{E_{\nu}}
\right) \label{elas1b}
\end{equation}
where in the mass basis $\mu_l^2$ takes the form
$$\mu^2_l=|c^2 \mu_{11}+s^2 \mu_{22}|^2$$
$$\mu^2_l=\mu^2_{21}~+~|c~\mu_{31}~+~s~exp(i \alpha_{CP})~\mu_{32}|^2$$
for Dirac and Majorana neutrinos respectively. For the definition of $c$ and $s$ see sec.
\ref{secno} below. In the case of Dirac neutrinos the off diagonal elements of the magnetic moment were meglected. The angle
$\alpha_{CP}$ is the relative CP phase of
of the dominant neutrino Majorana mass eigenstates present in the electronic neutrino.
 The contribution of the magnetic moment can also
be written as:
\begin{equation}
\left( \frac{d\sigma}{dT} \right)_{EM}=\sigma_0
 \left( \frac{\mu_l}{10^{-12}\mu_B} \right)^2 \frac{1}{T} \left( 1-\frac{T}{E_{\nu}} \right )
\label{elas2}
\end{equation}
The quantity $\sigma_0$ sets the scale for the cross section and
is quite small, $\sigma_0=2.5 \times 10^{-25}b$.

 The electron energy depends on the neutrino energy and the
scattering angle and is given by:
%$$T=\frac{X^2}{2 m_e}~~,~~X=2E_{\nu} \frac{m_e(m_e+E_{\nu})\cos{\theta}}
%{(m_e+E_\nu)^2-(E_{\nu} \cos{\theta})^2}$$
%The last equation can be simplified as follows:
$$T \approx \frac{ 2(E_\nu \cos{\theta})^2}{m_e}$$
% The electron
%energy depends on the neutrino spectrum.
For $E_{\nu}=18.6~KeV$
one finds that the maximum electron kinetic energy approximately
is \cite{GIOMATAR}:
$$ T_{max}=1.27~KeV$$
 Integrating the differential cross section between $0.1$ and $1.27~KeV$ we find that the total
cross section is:
$$\sigma=2.5~\sigma_0$$
It is tempting for comparison to express the above EM differential
cross section in terms of the weak interaction, near the threshold of
$0.1KeV$, as follows:
%\end{document}

\begin{equation}
\left( \frac{d\sigma}{dT} \right)_{EM}=\xi^2_1
\left( \frac{d\sigma}{dT} \right)_{Weak}
 \left( \frac{\mu_l}{10^{-12}\mu_B} \right)^2
\frac{0.1KeV}{T}  \left( 1-\frac{T}{E_{\nu}} \right)
\label{elas3}
\end{equation}
The parameter $\xi_1$ essentially gives the ratio of the
interaction due to the magnetic moment divided by that of the weak
interaction. Evaluated  at the energy of $0.1 KeV$ it becomes:
$$\xi_1 \approx 0.50 $$ Its value, of course, will be  larger if the magnetic
moment is larger than $10^{-12} \mu_B$.
Anyway the magnetic moment at these low energies can make a detectable
contribution provided that it is not much smaller than
 $10^{-12} \mu_B$.
 In many cases one would like to know the difference between the
 cross section of the electronic neutrino and that of one of the
 other flavors, i.e.
 \beq
 \chi(E_{\nu},T)=\frac{(d\sigma(\nu_e,e^-))/dT-d(\sigma(\nu_{\alpha},e^-))/dT}
 {d(\sigma(\nu_e,e^-))/dT}
 \label{elas4}
 \eeq
with $\nu_{\alpha}$  is either $\nu_{\mu}$ or $\nu_{\tau}$). Then
from the above expression for the differential cross-section one
finds:
%\begin{eqnarray}
\beq
 \chi=2
[\frac{2-(m_eT/E^2_{\nu})}{ff1(\theta_W)
 +2sin^2\theta_W(1-T/E_{\nu})^2-ff2(\theta_W)(m_eT/E^2_{\nu}}
 \label{elas5}
 \eeq
with
$$ff1(\theta_W)=(1+2\sin^2\theta_W)^2/(2sin^2\theta_W)~~,~~
ff2(\theta_W)= (1+2\sin^2\theta_W)$$
 For antineutrinos the above equation is slightly modified to yield
\beq
 \chi=2
[\frac{2-(m_eT/E^2_{\nu})}{
 2sin^2\theta_W+ff1(\theta_w)(1-T/E_{\nu})^2-ff2(\theta_W)(m_eT/E^2_{\nu}}
 \label{elas6}
\eeq
 Specializing  Eq. \ref{elasw} in the case of the antineutrino-electron
 scattering we get:
 \begin{eqnarray}
 \left(\frac{d\sigma}{dT}\right)_{weak}&=&\frac{G^2_F m_e}{2 \pi}\\
\nonumber
& & [ (2sin^2\theta_W)^2 +(1+2\sin^2\theta_W)^2 (1-T/E_{\nu})^2\\
 \nonumber
 &-&2sin^2\theta_W(1+2\sin^2\theta_W)(m_eT/E^2_{\nu})]
 \label{elaswb}
  \end{eqnarray}

 This last equation can be used to measure $\sin^2\theta_W$ at very
 low momentum transfers, almost 30 years after the first historic
 measurement by Reiness, Gur and Sobel  \cite{REINES}. In the present
 experiment we will measure the differential cross section as a
 function of $T$, which is a essentially a straight line. With
 sufficient statistics we expect to construct the straight line
 sufficiently accurately, so that we can extract $\sin^2\theta_W$
 both from the slope and the intercept achieving high precision.
 We should mention that the present method does not suffer from the
 well known supression of the weak charge associated with other low
 energy processes \cite{PRD} including the atomic physics experiments
 \cite{BENNET,BOUCHIAT}. This is due to the
 fact that the dependence on the Weinberg angle in these experiments
 comes from the
 neutral current vector coupling of the  electron and/or the proton,
 involving the combination $1-4\sin^2\theta_W \approx 0.1$. Thus in
 our approach it may not be necessary to go through an elaborate
 scheme of radiative corrections (see the recent work by Erler {\it et al}
  \cite{ERLER} and references therein).

\section{Experimental considerations}
\label{secec}
 In this section we will focus on the experimental
considerations

\begin{figure}
\hspace*{-0.0cm}
\includegraphics[height=.3\textheight]{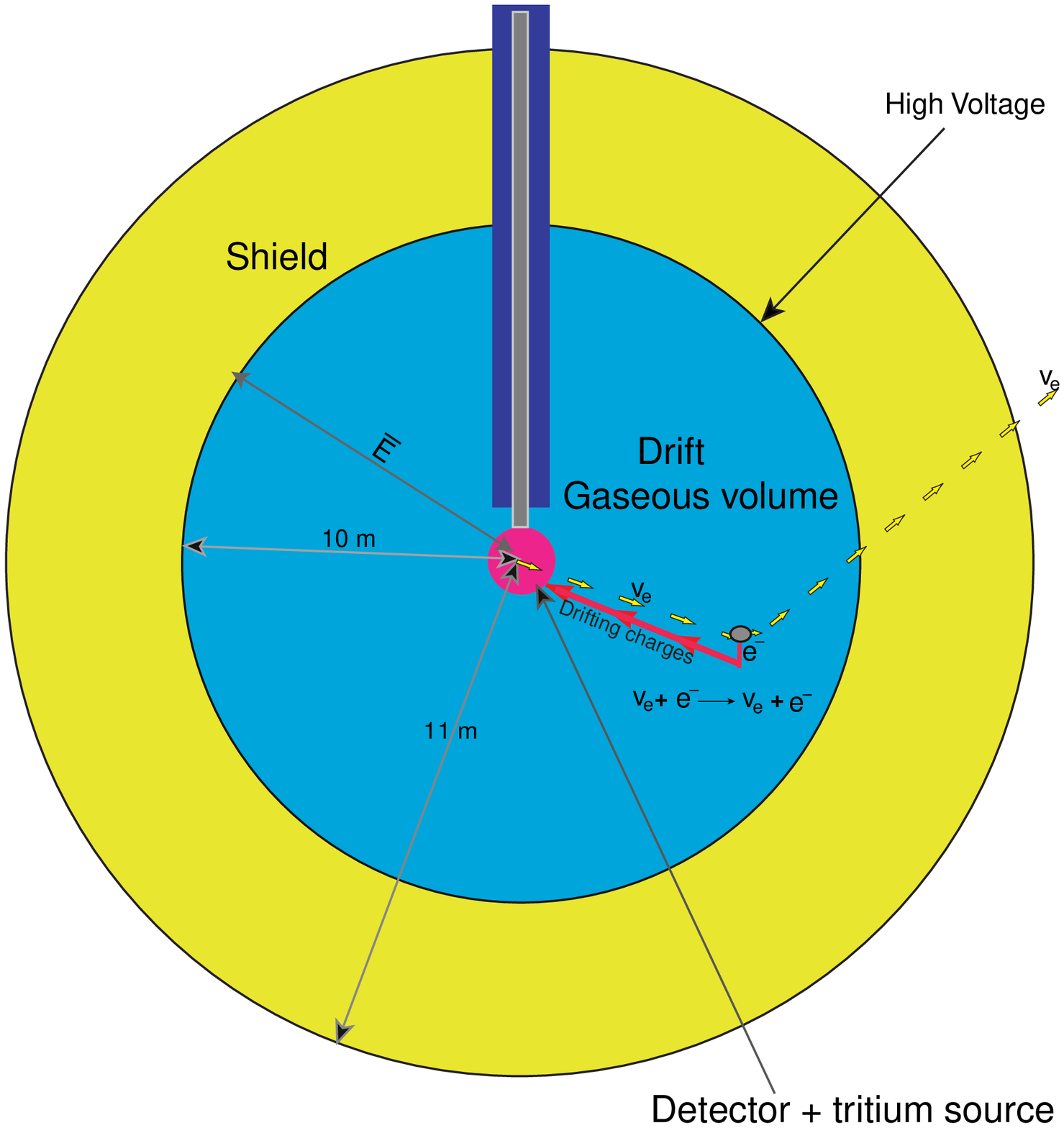}
\includegraphics[height=.35\textheight]{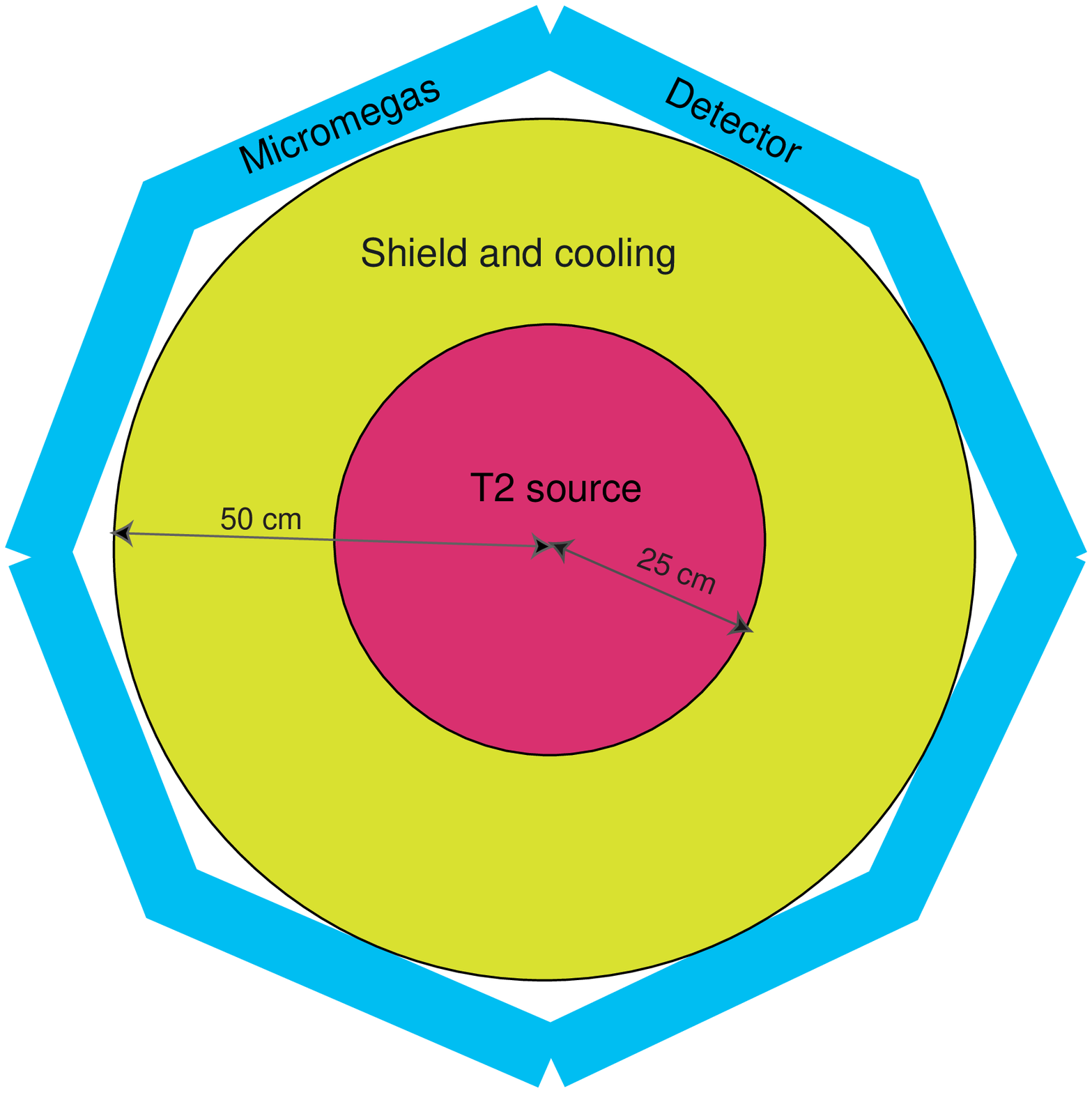}
\caption{ The principal features of the proposed TPC (a) and a
schematic view of the inner part of the vessel with the detector
and the tritium source (right).
 \label{draw1} }
\end{figure}

One of the attractive features of the gaseous TPC is its ability
to precisely reconstruct particle trajectories without precedent
in the  redundancy of experimental points, i.e. a bubble chamber
quality with higher accuracy and real time recording of the
events. Many proposals are actually under investigation to exploit
the TPC advantages for various astroparticle projects and
especially solar or reactor neutrino detection and dark matter
search \cite{GORO}-\cite{SNOWDEN}. A common goal is to fully
reconstruct the direction of the recoil particle trajectory, which
together with energy determination provide a valuable piece of
information. The virtue of using the TPC concept in such
investigations has been now widely recognized and a special
International Workshop has been recently organized in Paris
\cite{PARIS}. The study of low energy elastic neutrino-electron
scattering using a strong tritium source was envisaged in by
Bouchez and Giomataris \cite{BG} employing a large volume gaseous
cylindrical TPC. We will present here an alternate detector
concept with different experimental strategy based on a spherical
TPC design. A sketch of the principal features of the proposed TPC
is shown in Fig. \ref{draw1}.
%%%%%%%%%%%%%%%%%%%%%%%%%%%%%%%%%%%%%%%%%%%%%%%%%%%%%%%%%%%%%%%%%%%%%
%\begin{figure}
%\hspace*{-0.0 cm}
%\includegraphics[height=.3\textheight]{Detector+T2.eps}
%\caption{ A schematic view of the inner part of the vessel with
%the detector and the tritium source.
% \label{draw2} }
%\end{figure}
%%%%%%%%%%%%%%%%%%%%%%%%%%%%%%%%%%%%%%%%%%%%%%%%%%%%%%%%%%%%%%%%%%%%%
For a more detailed description of the apparatus, including the
neutrino source, the gas vessel and detailed study of the detector
\cite{ZIOUTAS,AALSETH}, \cite{MAGNON,ADRIAN} including a
discussion of  MICROMEGAS (MICROMEsh GAseous Structure)
\cite{GIOMA98,DERRE99,DERRE01} the reader is referred to our
previous work \cite{GIOMVER03}.

Our approach is radically different from all other neutrino
oscillation experiments in that it measures the neutrino
interactions, as a function of the distance source-interaction
point, with an oscillation length that is fully contained in the
detector; it is equivalent to many experiments made in the
conventional way where the neutrino flux is measured in a single
space point.
  Furthermore, since the oscillation length is comparable to the
  detector depth, we expect an
exceptional signature: a counting rate oscillating from the triton
source location to the depth of the gas volume, i.e. at first a
decrease, then a minimum and finally an increase. In other words
we will have a full observation of the oscillation process as it
has already been done in accelerator experiments with neutral
strange particles ($K^0$).

To summarize:
\begin{itemize}
\item The aim of the proposed detector will be the detection of very low
 energy neutrinos emitted by a strong tritium source through their elastic
 scattering on electrons of the target.
\item The $(\nu,e)$ elastic differential cross section is the sum of the
 charged and neutral current contributions (see sec. \ref{secens}) and
is a function of the energy. It is, however,  it is quite small,
 see Eq. (\ref{weekval}).
\item Integrating this cross section up to energies of $15~KeV$ we get a very
 small value, $\sigma=0.4\times 10^{-47}cm^2$. This means that, to get a
significant signal in the detector, for 200 Mcurie
tritium source (see next section) we will need about 20 kilotons of gaseous
 material.
\item The elastic $(\nu,e)$ cross section, being dominated by the charged
 current, especially for low energy electrons (see Fig. \ref{chi} below),
will be different from that of the other flavors, which is due to the
 neutral current alone. This will allow us to observe neutrino oscillation
enabling a modulation on the counting rate along the oscillation length.
The effect depends on the electron energy T as is shown in sec. \ref{secno}
\end{itemize}
We assume a spherical type detector, described in the previous
section, filed with Xenon gas at NTP and a tritium source of 20
kg, providing a very-high intensity neutrino emission of $6
\times10^{18} /s$. The Monte Carlo program is simulating all the
relevant processes:
\begin{itemize}
\item Beta decay and neutrino energy random generation
\item Oscillation process of $\nu_e$ due to the small mixing $\theta_{13}$ (see
Eq. \ref{osceq} below).
\item Neutrino elastic scattering with electrons of the gas target
\item Energy deposition, ionization processes and transport of charges to the
Micromegas detector.
\end{itemize}

\begin{figure}
\hspace*{-0.0 cm}
\includegraphics[height=.4\textheight]{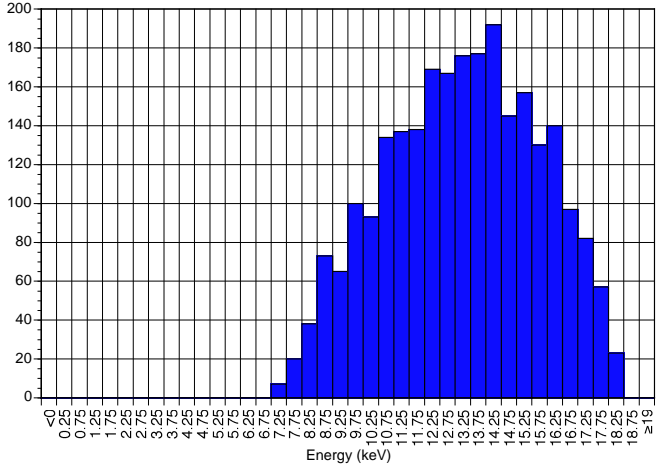}
\caption{ Neutrino energy distribution with an imposed energy
cutoff of $200~eV$.
 \label{ioa3} }
\end{figure}

 First Monte
Carlo simulate are giving a resolution of better than 10 cm, which
is good enough for our need. In Fig. \ref{ioa3} the energy
distribution of the detected neutrinos, assuming a detection
threshold of 200 eV, is exhibited. The energy is concentrated
around 13 keV with a small tail to lower values.
%%%%%%%%%%%%%%%%%%%%%%%%%%%%%%%%%%%%%%%%%%%%%%%%%%%%%%%%%%%%%%%%%%%%%

In Figs \ref {osc} and \ref{osc2} below we  show the number of
detected elastic events as function of the distance L in bins of
one meter for several hypothesis for the value of the mixing angle
$\theta^2_{13}= 0.170, 0.085$ and $0.045$.
 We observe a decreasing of the signal up to about
6.5 m and then a rise. Backgrounds are not yet included in this
simulation but the result looks quite promising; even in the case
of the lowest mixing angle the oscillation is seen, despite
statistical fluctuations. We should point out that in the context
of this experiment complete elimination of the backgrounds is not
necessary. It is worth noting that:
\begin{itemize}
\item A source-off measurement at the beginning of the experiment will yield the
background level to be subtracted from the signal.
\item Fitting the observed oscillation pattern will provide, for the first time, a stand
alone measurement of the oscillation parameters, the mixing angle
and the square mass square difference.
\item Systematic effects due to backgrounds or to bad estimates of the neutrino
flux, which is the main worry in most of the neutrino experiments,
are highly reduced in this experiment.
\end{itemize}
\section{ A simple phenomenological neutrino mixing matrix-
Simple expressions for neutrino oscillations}
\label{secno}
 The available neutrino oscillation data (solar \cite{SOLAROSC} and
  atmospheric \cite{SUPERKAMIOKANDE})as well as the
 KamLAND \cite{KAMLAND} results can
adequately be described by the following matrix:

$\left ( \begin{array}{c}\nu_e \\
 \nu_{\mu} \\
\nu_{\tau} \end{array} \right )=
\left ( \begin{array}{ccc}
c&s&\delta\\
-\frac{s+c \delta}{\sqrt 2}&  \frac{c-s \delta}{\sqrt 2}& \frac{1}{\sqrt 2}\\
 \frac{s-c \delta}{\sqrt 2}& -\frac{c+s \delta}{\sqrt 2}& \frac{1}{\sqrt 2}\\
 \end{array} \right )=
\left ( \begin{array}{c}\nu_1 \\
\nu_2 \\
\nu_3 \end{array} \right )$\\

Up to order $\delta^2$ ($\delta^2=4 \times 10^{-2}$). Sometimes we
will use $\theta_{13}$ instead of $\delta$. Knowledge of this
angle is very crucial for CP violation in the leptonic sector,
since it may complex even if the neutrinos are Dirac particles. In
the above expressions we have not absorbed the phases arising, if
the neutrinos happen to be Majorana particles,$\nu_k \xi_k = C
~{\overline{\nu}}_k^T$ where C denotes the charge conjugation,
 $\xi_k=e^{i\lambda_k}$, which guarantee that the eigenmasses are positive.
The other entries are:
$$c=\cos \theta_{solar}~,~s=\sin \theta_{solar}$$
determined from the solar neutrino data \cite{SOLAROSC},
\cite{BAHCALL02}-\cite{BARGER02}
$$\tan^2 \theta_{solar} \approx  0.35-0.42$$
$$0.26 \le \tan^2 \theta_{solar} \le 0.85~~(3\sigma)$$
while the analysis of KamLAND results
\cite{BAHCALL02}-\cite{BARGER02} yields:
$$\tan^2 \theta_{solar} \approx  0.46-0.64$$
$$0.29 \le \tan^2 \theta_{solar} \le0.86~~(3\sigma)$$

%\section{ Simple expressions for neutrino oscillations}
%\label{secno}
%\subsection{In the absence of magnetic field}
\begin{itemize}
\item Solar neutrino Oscillation (LMA solution) is given by:
$$P(\nu_e \rightarrow \nu_e)\approx 1-(\sin 2 \theta_{solar})^2 \sin^2(\pi \frac{L}{L_{21}})$$
$$L_{21}=\frac{4 \pi E_{\nu}}{\Delta m^2_{21}}$$
The analysis of both the neutrino oscillation experiments as well
as KamLAND  \cite{BAHCALL02}-\cite{BARGER02} yield
$$\Delta
m^2_{21}=|m_2^2-m_1^2|=(5.0-7.5)\times 10^{-5}(eV)^2$$
\item The
Atmospheric Neutrino Oscillation takes the form:
$$P(\nu_{\mu} \rightarrow \nu_{\tau})= \sin^2(\pi \frac{L}{L_{32}})$$
$$L_{32}=\frac{4 \pi E_{\nu}}{\Delta m^2_{32}} \rightarrow \Delta m^2_{32}=|m_3^2-m_2^2|=2.5\times 10^{-3}(eV)^2$$
\item We conventionally write
$$\Delta m^2_{32}=\Delta m_{atm}^2~~,~~\Delta m^2_{21}=\Delta m_{sol}^2$$
\item Corrections to disappearance experiments
\beq P(\nu_e \rightarrow \nu_e)= 1-\frac{(\sin 2 \theta_{solar})^2
\sin^2(\pi \frac{L}{L_{21}}) +4 \delta^2 \sin^2(\pi
\frac{L}{L_{32}})}{(1+\delta^2)^2}
 \label{osceq}
 \eeq
 \item The probability for $\nu_e\rightarrow\nu_{\mu}$ oscillation
 takes the form:
\begin{eqnarray}
 P(\nu_e \rightarrow \nu_{\mu})&=& \frac{ \left [(\sin 2
\theta_{solar})^2 +\delta \sin{4\theta_{solar}}\right] \sin^2(\pi
\frac{L}{L_{21}})}{(1+\delta^2)^2}\\
\nonumber
&+& \frac{4 \delta^2 \sin^2(\pi \frac{L}{L_{32}})}{(1+\delta^2)^2}
 \label{osceqa}
\end{eqnarray}
\item While the oscillation probability  $\nu_e\rightarrow\nu_{\tau}$ becomes:
\begin{eqnarray}
 P(\nu_e \rightarrow \nu_{\tau})&=& \frac{ \left [(\sin 2
\theta_{solar})^2 -\delta \sin{4\theta_{solar}}\right] \sin^2(\pi
\frac{L}{L_{21}})}{(1+\delta^2)^2}\\
\nonumber
&+& \frac{4 \delta^2 \sin^2(\pi \frac{L}{L_{32}})}{(1+\delta^2)^2}
 \label{osceqb}
\end{eqnarray}
 \end{itemize}
  From the above expression we see that the small
amplitude $\delta$ term dominates in the case of triton neutrinos
($L \le L_{32}~,L_{21}=50L_{32}$). In a different notation
$4 \delta^2  \approx sin^2 { 2 \theta_{13}}$

\begin{figure}
\hspace*{-0.0 cm}
\includegraphics[height=.15\textheight]{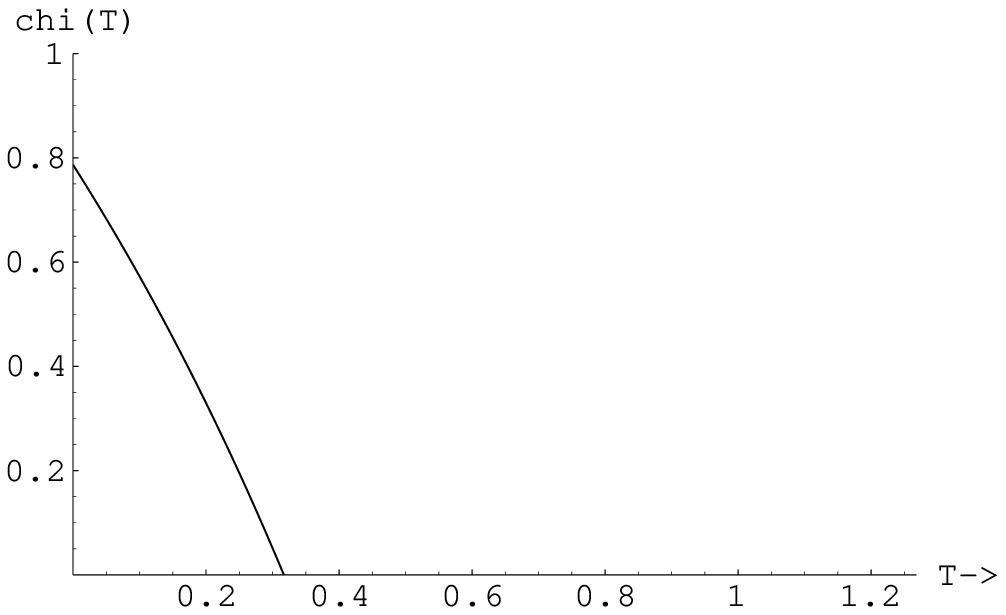}
\includegraphics[height=.15\textheight]{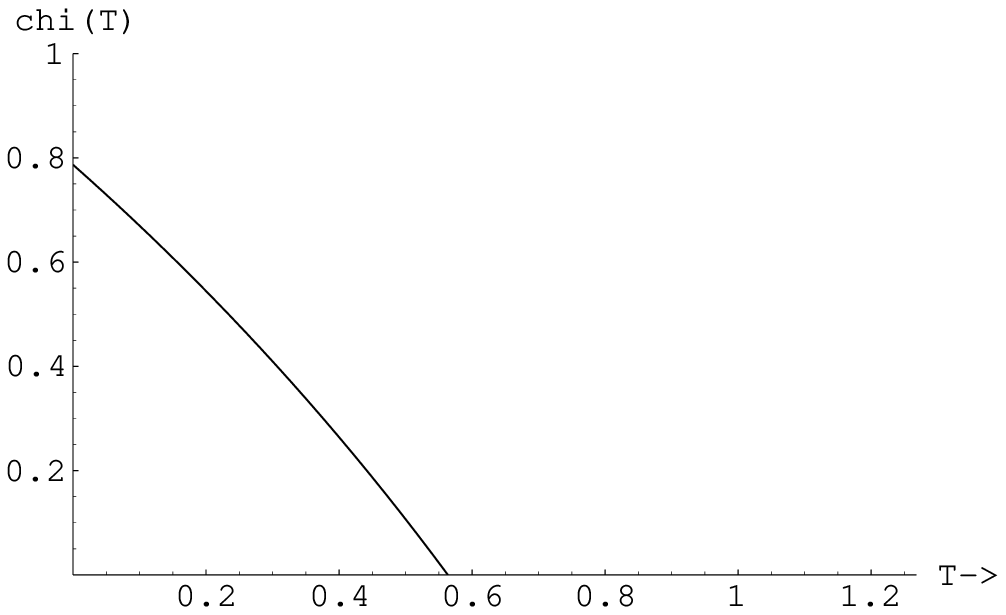}
\caption{ The parameter $\chi$ as a function of the electron
kinetic energy T for $E_{\nu}=9.0~KeV$ on the left and $12.0~KeV$
on the right.
 \label{chi1} }
\end{figure}

\begin{figure}
\hspace*{-0.0 cm}
\includegraphics[height=.15\textheight]{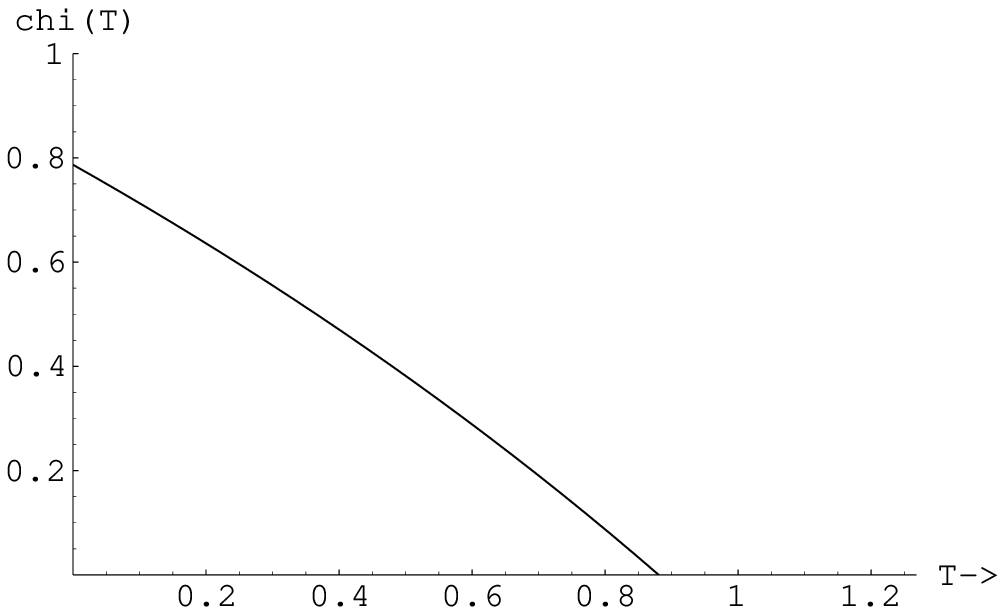}
\includegraphics[height=.15\textheight]{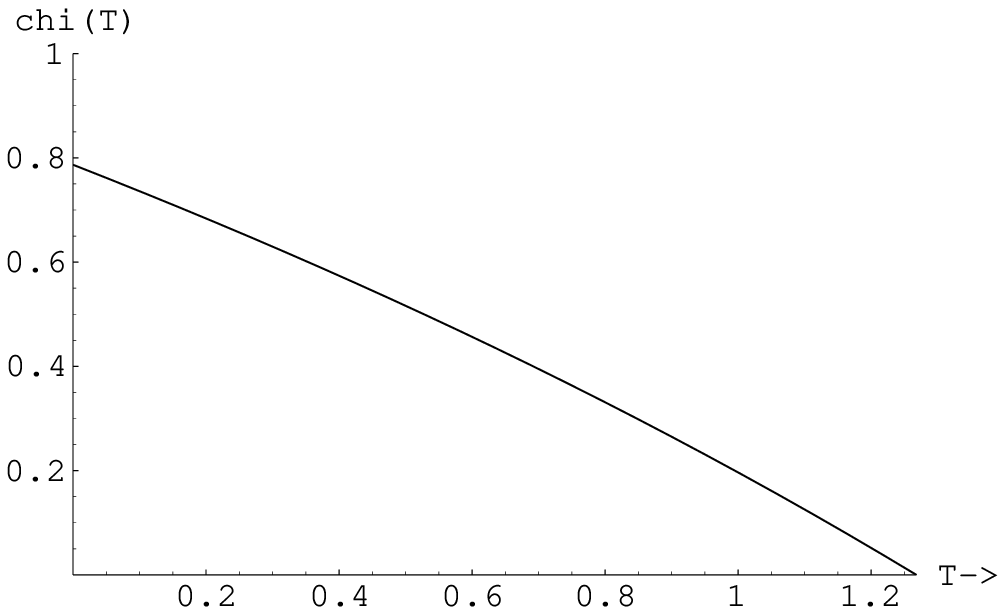}
\caption{ The parameter $\chi$ as a function of the electron
kinetic energy T for $E_{\nu}=15.0~KeV$ on the left and $18.0~KeV$
on the right.
 \label{chi2} }
\end{figure}

In the proposed experiment the neutrinos will be detected via the
 recoiling electrons. If the neutrino-electron cross section were the same for
 all neutrino species one would not observe any oscillation at
 all. We know, however, that the electron neutrinos behave very
 differently due to the charged current contribution, which is not
 present in the other neutrino flavors. Thus the number of the
 observed electron events ($ELEV$) will vary as a function of $L/E_{\nu}$ as
 follows:\\

\begin{eqnarray}
ELEV& & \propto \frac{d(\sigma(\nu_e,e^-))}{dT}\\
\nonumber
& & \left [1-\chi(E_{\nu},T) \frac{(\sin 2
\theta_{solar})^2 \sin^2(\pi \frac{L}{L_{21}}) +4 \delta^2
\sin^2(\pi \frac{L}{L_{32}})}{(1+\delta^2)^2} \right]\\
 \label{eventeq}
\end{eqnarray}
where
$$\chi(E_{\nu},T)=\frac{(d\sigma(\nu_e,e^-))/dT-d(\sigma(\nu_{\alpha},e^-))/dT}{d(\sigma(\nu_e,e^-))/dT}$$
($\nu_{\alpha}$  is either $\nu_{\mu}$ or $\nu_{\tau}$). In other
words $\chi$ represents the fraction of the $\nu_e$-electron
cross-section, $\sigma(\nu_e,e^-)$, which is not due to the
neutral current. Thus the apparent disappearance oscillation
probability will be quenched by this fraction. As we will see
below, see section \ref{secens}, the parameter $\chi$, for
$sin^2\theta_W=0.2319$, can be cast in  the form:
 \beq
\chi(E_{\nu},T)=\frac{2[2-(m_eT/E^2_{\nu})]}{4.6199+0.4638(1-T/E{\nu})^2-1.4638(m_eT/E^2_{\nu})}
 \label{chi}
 \eeq
 For antineutrinos the previous expression becomes:
  \beq
\chi(E_{\nu},T)=\frac{2[2-(m_eT/E^2_{\nu})]}{0.46384+4.6199(1-T/E{\nu})^2-1.4638(m_eT/E^2_{\nu})}
 \label{chib}
 \eeq
We thus see that the parameter $\chi$ depends not only on the
neutrino energy, but on the electron energy as well, see Figs
\ref{chi1}-\ref{chi2}.
Since in our experiment $T$ is very low there is no essential
difference between the two expressions for $\chi$.

%%%%%%%%%%%%%%%%%%%%%%%%%%%%%%%%%%%%%%%%%%%%%%%%%%%%%%%%%%%%%%%%%%%%%
%%%%%%%%%%%%%%%%%%%%%%%%%%%%%%%%%%%%%%%%%%%%%%%%%%%%%%%%%%%%%%%%%%%%%
%%%%%%%%%%%%%%%%%%%%%%%%%%%%%%%%%%%%%%%%%%%%%%%%%%%%%%%%%%%%%%%%%%%%%
It interesting to see that, for a given neutrino energy, $\chi$,
as a function of $T$, is almost a straight line. We notice that,
for large values of $T$, the factor $\chi$ is suppressed, which is
another way of saying that, in this regime, in the case of
$(\nu_e,e^-)$ differential cross-section the charged current
contribution is cancelled by that of the neutral current.

 In order
to simplify the analysis one may try to replace $\chi$ by an
average value $\bar{\chi}(E_{\nu})$, e.g. defined by:
 \beq
 \bar{\chi}(E_{\nu})=\frac{1}{T_{max}(E_{\nu})}\int_0^{T_{max}(E_{\nu})}
 \chi(E_{\nu},T)~dT
 \label{chiav}
 \eeq
 Then surprisingly one finds $\bar{\chi}(E_{\nu})$ is independent
 of $E_{\nu}$ with a constant value of $0.42$. This is perhaps a
 rather high price one may have to pay for detecting the neutrino
 oscillations as proposed in this work. One may turn this into an advantage,
however, since the disappearance dip in Eq. \ref{eventeq}, in
addition to its dependence on the familiar parameters , it also
depends on the electron energy.

\begin{figure}
\hspace*{-0.0 cm}
\includegraphics[height=.35\textheight]{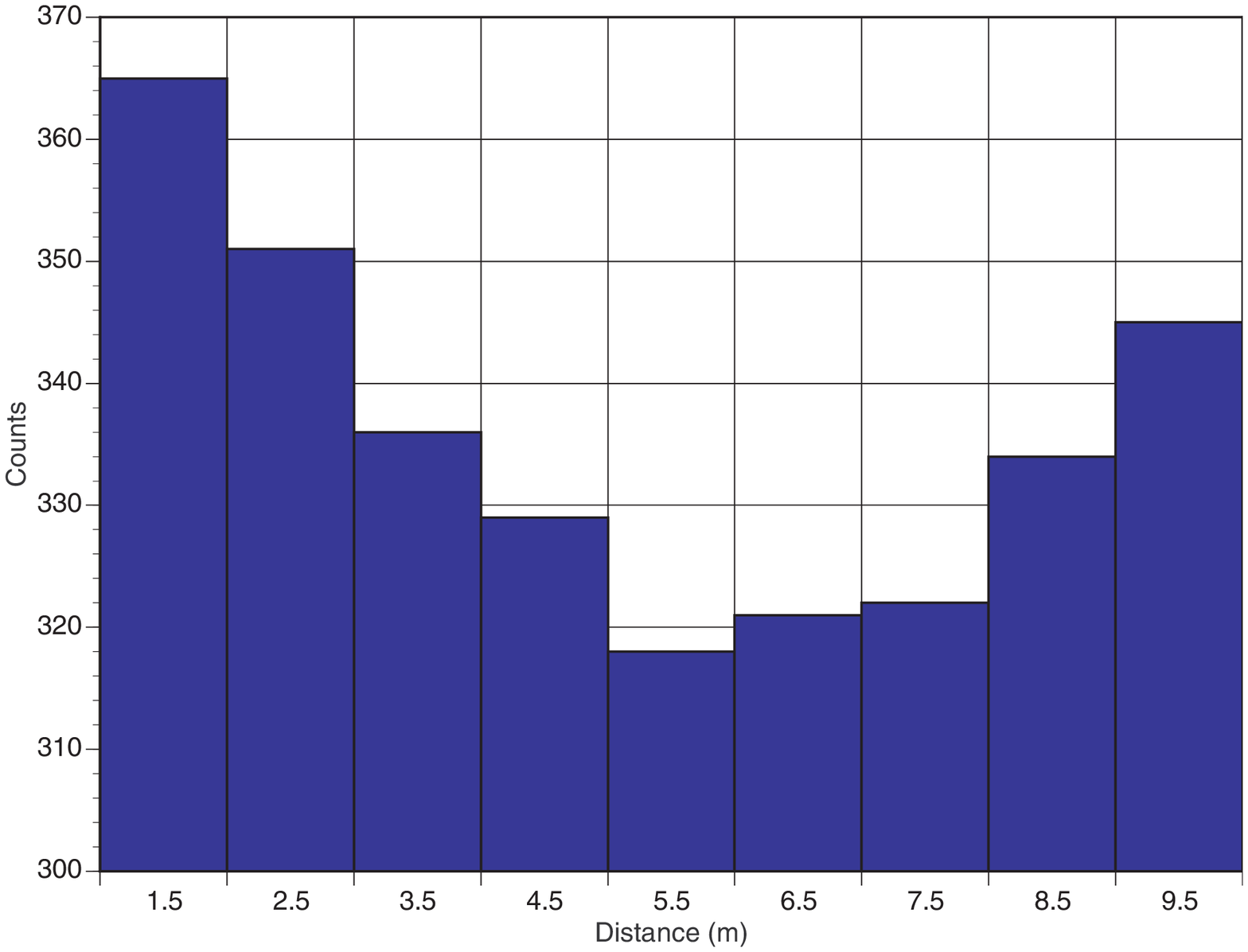}
\includegraphics[height=.35\textheight]{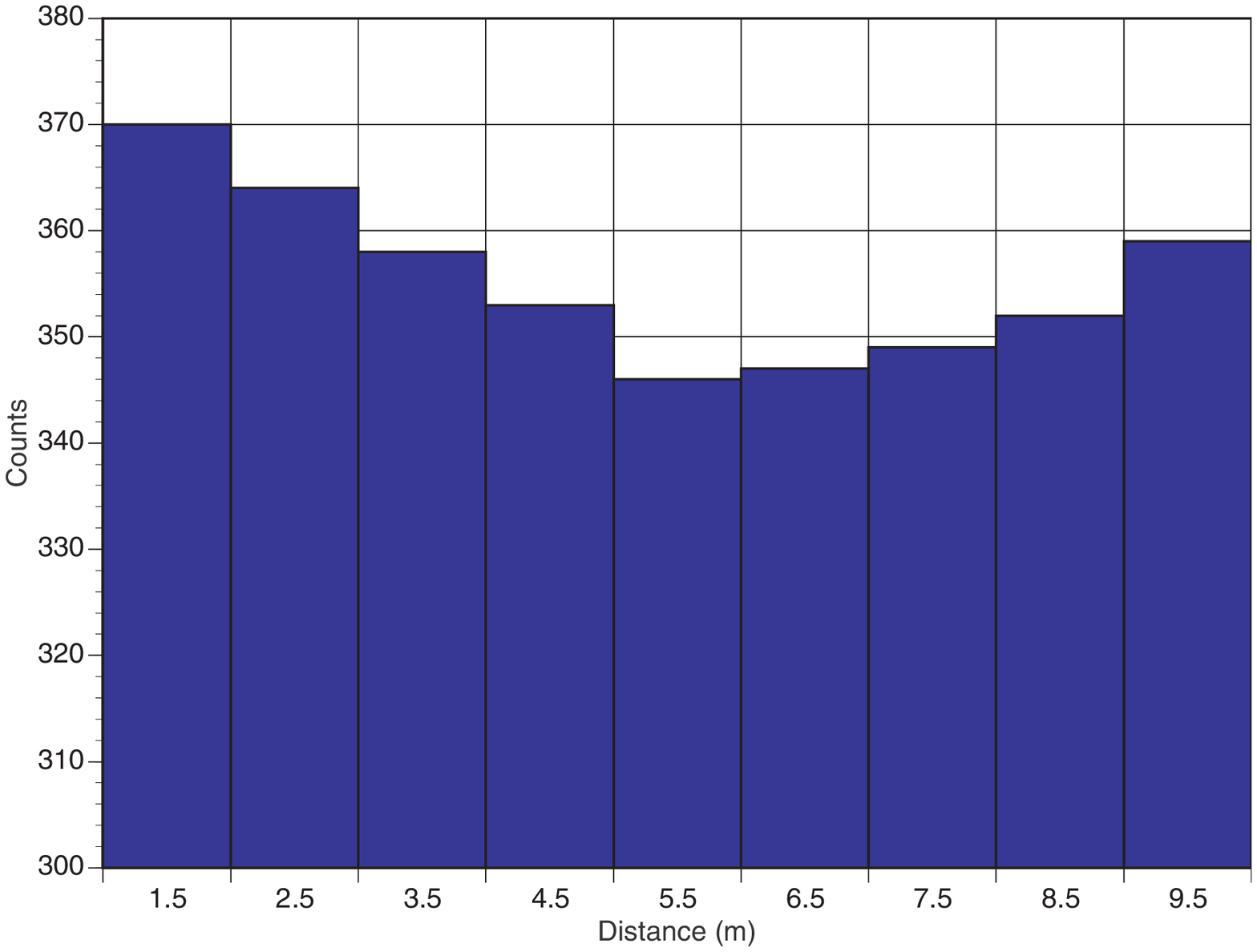}
\caption{Simulation of $\nu_e$ disappearance due to the large
 $\Delta m^2=2.5 \times 10^{-3} (eV)^2$
involving the small mixing angle  $\sin{2 \theta_{13}}^2$. The
parameter $\chi(E_{\nu},T)$ was not included in making the plots.
 On the left we show
results for  $\sin{2 \theta_{13}}^2=0.170$ , while on
 the right we show results for  $\sin{2 \theta_{13}}^2=0.085$.
  One expects to unambiguously see the full oscillation inside
the detector with the maximum disappearance occurring around
$6.5m$.
\label{osc}}
\end{figure}

\begin{figure}
\hspace*{-0.0 cm}
\includegraphics[height=.4\textheight]{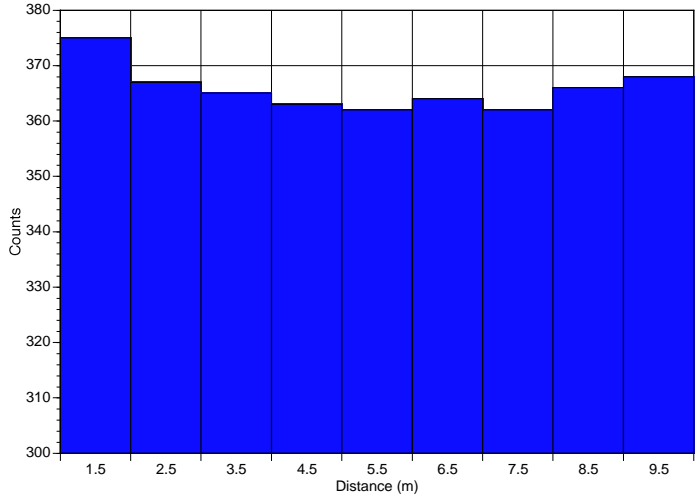}
\caption{ The same as in Fig. \ref{osc} for  $\sin{2
\theta_{13}}^2=0.045$. \label{osc2} }
\end{figure}

 Anyway in the experiment involving a triton target one will
actually observe a sinusoidal oscillation as a function of the
source-detector distance $L$ with an amplitude, which is
proportional to the square of the small mixing angle $\delta$. The
relevant oscillation length is given by:
$$L_{32}=2.476m \frac{E_{\nu} (MeV)}{\Delta m^2_{32}((eV)^2)}$$
In the present experiment for an average neutrino energy
 $E_{\nu} \approx 13KeV$ and $\Delta m^2_{32}=2.5 \times 10^{-3} (eV)^2$ we
find
$$L_{32} \approx 13.5m$$
In other words the maximum will occur close to the source at
about $L=7.5m$. Simulations of the above neutrino oscillation
involving $\nu_e$ disappearance due to the large $\Delta m^2=2.5
\times 10^{-3}$, i.e associated with the small mixing $\delta$,
are shown in Figs \ref{osc}- \ref{osc2}. One clearly sees that
the expected oscillation, present even for $\sin{2\theta_{13}}^2$.
 as low as $0.045$, will
occur well inside the detector.
%%%%%%%%%%%%%%%%%%%%%%%%%%%%%%%%%%%%%%%%%%%%%%%%%%%%%%%%%%%%%%%%%%%%%
%%%%%%%%%%%%%%%%%%%%%%%%%%%%%%%%%%%%%%%%%%%%%%%%%%%%%%%%%%%%%%%%%%%%%

 Superimposed on this oscillation one will
see an effect due to the smaller mass difference, which will
increase quadratically with the distance $L$.

\begin{figure} \hspace*{-0.0 cm}
\includegraphics[height=.15\textheight]{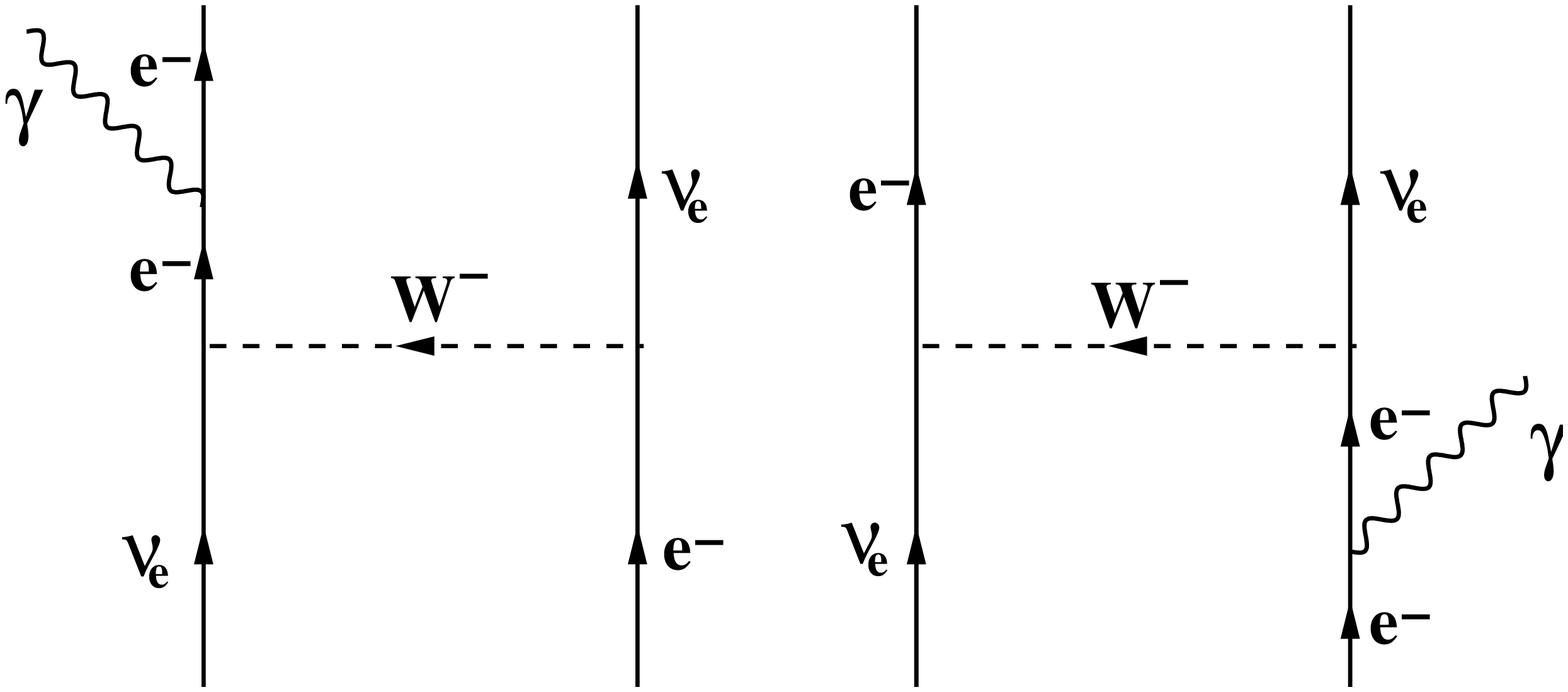}
\includegraphics[height=.15\textheight]{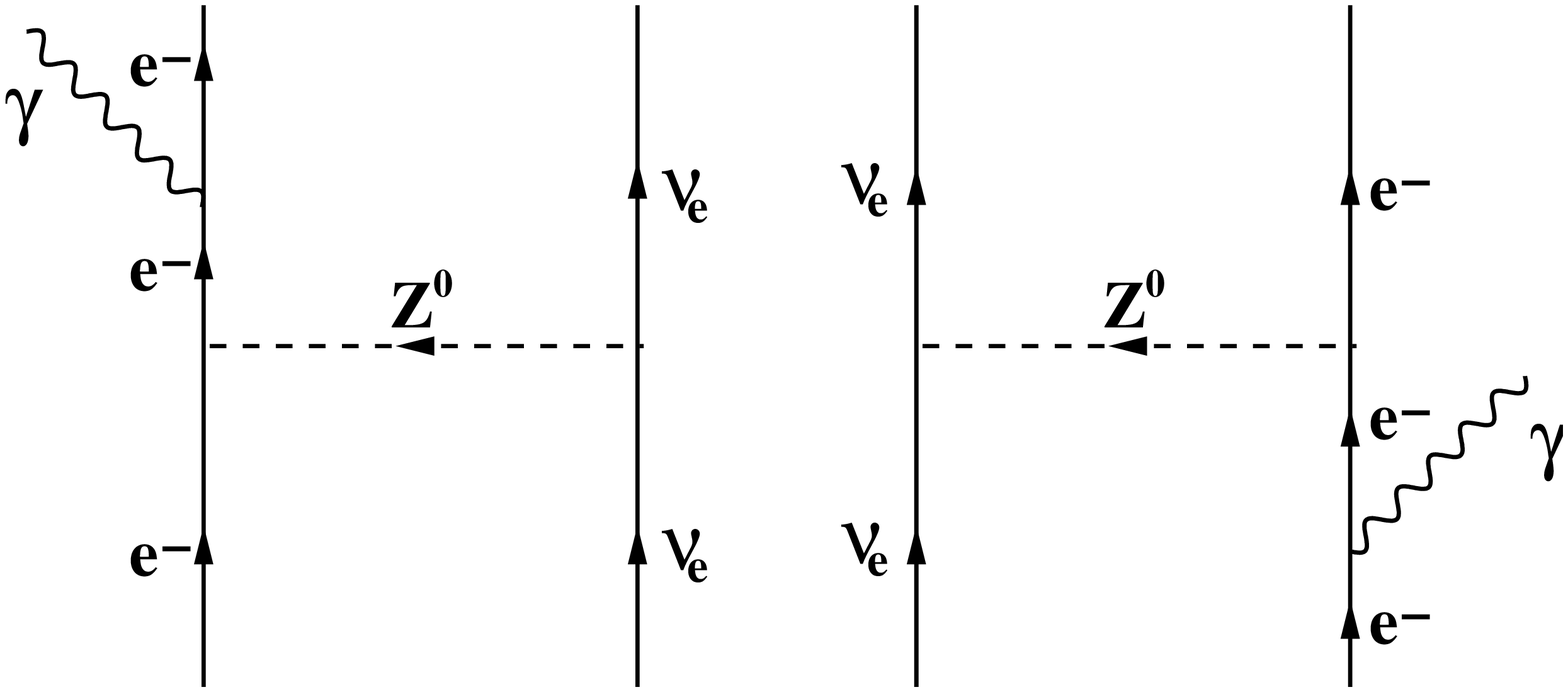}
\caption{ The Feynman diagrams contributing to radiative neutrino
electron scattering via the charged current (left) and neutral
current (right).
 \label{charneut} }
\end{figure}

 The above simple neutrino oscillation formulas get modified i) In the presence
 of a magnetic field if the neutrino has a magnetic moment and/or ii) If
 the heavier neutrinos have a finite life time.
 section{Radiative neutrino-electron scattering}
\label{secrens}
  The radiative neutrino decay for low energy neutrinos is perhaps
unobservable. Radiative neutrino decay in the presence of matter,
in our case electrons, is however observable.
$$\nu_e(p_{\nu})+e^- \longrightarrow \nu_e (p^{'}_{\nu})+e^-(p_e)+\gamma(k)$$
 This occurs  via the collaborative
effect of electromagnetic and weak interactions as is shown in
Fig. \ref{charneut}.
%%%%%%%%%%%%%%%%%%%%%%%%%%%%%%%%%%%%%%%%%%%%%%%%%%%%%%%%%%%%%%%%%%%%%
%\begin{figure}
%\hspace*{-0.0 cm}
%\includegraphics[height=.3\textheight]{neutral.eps}
%\caption{ The Feynman diagrams contributing to radiative neutrino
%electron scattering via the neutral current. \label{neutral} }
%\end{figure}
%%%%%%%%%%%%%%%%%%%%%%%%%%%%%%%%%%%%%%%%%%%%%%%%%%%%%%%%%%%%%%%%%%%%%

 The evaluation of the cross section
associated with these diagrams is rather complicated, but in the
present case the electrons are extremely non relativistic. Thus in
intermediate electron propagator we can
retain the mass rather than the momenta by writing
 (see Fig. \ref{charneut})
 $(p_i-k)^{\mu}\gamma_{\mu}+m_e\approx2m_e$,
 $(p_f+k)^{\mu}\gamma_{\mu}+m_e\approx2m_e$
 (exact results without
this approximation will appear elsewhere). Then after some
tedious, but straight-forward, trace calculations one can perform
the angular integrals over the three-body final states. Thus  to
leading order in the electron energy one gets:
\begin{eqnarray}
k\frac{d\sigma(k,T_e)}{dT_e~dk}&=&\sigma_{\gamma}\frac{m_e}{2\bar{E}^2_{\nu}}
[(1-\frac{k}{E_{\nu}})g_V^2\\
\nonumber
&+& \left(\frac{E^2_{\nu}}{m_e^2}(1-\frac{k}{E_{\nu}})-\frac{1}{8}(4g^2_V+g_Vg_A)\right)
\frac{T_e~m_e}{E^2_{\nu}}]
 \label{radia1}
\end{eqnarray}
 where $\bar{E}_{\nu}=13.0~KeV$ is the average neutrino energy and
$$\sigma_{\gamma}=4\frac{8}{\pi^2}(G_F~\bar{E}_{\nu})^2~\alpha
\approx2.0\times 10^{-13}pb$$ sets the scale for this process.
This momentum depends on the photon momentum k and the scattering
angles. For a given $k$ is restricted as follows:
$$0 \leq T_e \leq \frac{2(E_{\nu}-k)^2}{m_e}$$
From the above equations we cam immediately see that this process
is roughly of order $\alpha$ down compared to the weak
neutrino-electron scattering cross-section. We also notice that
the total cross section diverges logarithmically as the photon
momentum goes to zero, reminiscent of the infrared divergence of
Bremsstrahlung radiation. In our case we will adopt a lower photon
momentum cutoff as imposed by our detector. We also notice that
$\sigma_{\gamma}$, characterizing  this process, is only a factor
of three smaller than $\sigma_0$ characterizing the neutrino
electron scattering cross section due to the magnetic moment. We
should bare in mind, however, that:
\begin{enumerate}
\item The  magnetic moment is not known. $\sigma_0$ was obtained
with the rather optimistic value $\mu_{\nu}=10^{-12}\mu_{B}$,
which is two orders of magnitude smaller than the present
experimental limit.
\item One now has the advantage of observing not only the electron but the photon as
well.
 \end{enumerate}
Integrating over the electron energy we get:
\begin{eqnarray}
k\frac{d\sigma(k}{dk}&=&\sigma_{\gamma}\frac{E^2_{\nu}}{\bar{E}^2_{\nu}}
(1-\frac{k}{E_{\nu}})^3 [ g_V^2+
(\frac{E^2_{\nu}}{m_e^2}(1-\frac{k}{E_{\nu}})\\
\nonumber
&-&\frac{1}{8}(4g^2_V+g_V~g_A))
(1-\frac{k}{E_{\nu}}) ]
 \label{radia2}
\end{eqnarray}
 Integrating this cross-section with respect
to the photon momentum we get:
 \beq
\sigma_{total}=\sigma_{\gamma}~\frac{E_{\nu}^2}{\bar{E}_{\nu}^2}
\left[(\frac{1}{2}g_V^2-\frac{1}{4}g_A^2) \ln
\frac{E_{\nu}}{E_{cutoff}}+ \frac{25}{24}g_V^2-\frac{5}{48}g_A^2
\right]
 \label{radia3}
  \eeq
  with the energy cutoff $E_{cutoff}$ determined by the detector.

 We have considered in our discussion only electron targets. For such low
 energy antineutrinos the charged current cannot operate on hadronic targets,
  since this process is not allowed so long as the target, being stable,
  is not capable of undergoing positron decay. The
  neutral current, however, can always make a contribution.
 % $E_{cutoff}< 0.223E_{\nu}~$.
\section {Summary and outlook}
The perspective of the experiment is to provide high statistics
-redundant, high precision measurement and minimize as much as
possible the systematic uncertainties of experimental origin,
which could be the main worry in the results of existing
experiments. The physics goals of the new atmospheric neutrino
measurement are summarized as follows:
\begin{enumerate}
\item Establish the phenomenon of neutrino oscillations with a
different experimental technique free of systematic biases. The
oscillation length, associated with the small mixing angle
$\sin\theta_{13}$ in the electronic neutrino, is fully contained
in our detector. Thus one hopes to measure all the oscillation
parameters, including the small mixing angle, clarifying this way
the nature of the oscillation mechanism. \item A high sensitivity
measurement  of the neutrino magnetic moment, via electron
neutrino scattering. At the same time radiative electron neutrino
scattering will be investigated, exploiting the additional photon
signature. \item A precise measurement of $\sin^2\theta_W$ at very
low momentum transfer, difficult to achieve in other experiments.
\item A new experimental investigation of neutrino decay. \item
Other novel improvements of the experimental sensitivity are
possible and must be investigated. The benefit of increasing the
gas pressure of the detector, which leads to a proportional
increase in the number of events, must be investigated. \item The
estimates presented above correspond to a year of data taking. In
our experiment, however, in addition to increasing the pressure,
there is no problem in increasing the data taking period up to
 10 years or even longer, increasing our statistics accordingly.
 Thus the prospect of reaching 100000 detected events is quite
realistic. This significant increase of the event rate is
definitely going to be a great step forward towards improving the
experimental accuracy and reducing the impact of background
uncertainties.
\end{enumerate}

\end{document}